\documentclass[prb,preprint]{revtex4-1} % style for Physical Review B and AJP are similar

\usepackage{amsmath} % need for subequations
\usepackage{color}
\usepackage{graphicx} % for figures
\begin{document}
\textit{The following article has been accepted by the American Journal of Physics. After it is published, it will be found at http://scitation.aip.org/ajp/.}

\title{On the unavoidability of the interpretations of Quantum Mechanics}
%Lines break automatically or can be forced with \\
\author{Roberto Beneduci}
\email{rbeneduci@unical.it}
%\altaffiliation[Also at ]{home.} % optional
\affiliation{ Department of Physics University of Calabria (Cosenza, Italy)\\
and\\
Istituto Nazionale di Fisica Nucleare, gruppo collegato Cosenza (Cosenza, Italy)
}
 %optional
\author{Franklin E. Schroeck, Jr.}
\email{fschroec@du.edu}
\affiliation{University of Denver (Colorado, USA)\\
and\\
Florida Atlantic University (Florida, USA)}

\date{\today}

%\begin{abstract}
%We give some examples of the use of \LaTeX\ that we hope will be
%helpful in preparing manuscripts for \ajp.
%\end{abstract}

\maketitle

\section{Introduction}\label{intr}
\noindent
A recent letter\cite{Kampen} by N.~G.~van Kampen stimulated an interesting debate\cite{Hobson,Henry,Hobson1} on the interpretation of quantum mechanics. The central theme of the debate was a criticism to the (also recent) presence in the literature of ``voluminous discussions'' about the interpretation of quantum mechanics. That there still exists such discussions is what van Kampen calls the ``scandal of quantum mechanics.'' 

We claim that a weak point in the debate was a lack of a definition of the term \textit{interpretation\,}.  In the present note, we would like to make that meaning precise and to show how such a clarification is necessary in order to avoid misunderstandings. The concept of interpretation plays a key role in the history of physics, so we hope the present analysis might be helpful to the student.  

Max Jammer,\cite{Jammer} in his famous book \textit{The philosophy of quantum mechanics}, distinguishes four different meanings of the word interpretation.  In what follows we restrict our attention to only three different meanings, which we now try to clarify.

%%%%%%%%%%%%%%%%%%%%%%%%%%%%%%%%%%%%%%%%
\section{Statistical regularities: The minimal instrumentalist interpretation}\label{I}

A physical theory $T$ has at least two components: \cite{Suppe} 1) the formalism $F$, or mathematical apparatus, of the theory, and 2) the rules of correspondence $R$ that establish a link between the formalism and the results of measurement. 
As an example, in the formalism of quantum mechanics based on a Hilbert space $\cal{H}$, the formalism has at least two components: states represented by density operators $\rho$ on $\cal{H}$, and observables represented by self-adjoint operators ${A}$, on $\cal{H}$. To each vector $\psi\in \cal{H}$ with $\|\psi\|=1$, there corresponds a pure state $\rho=\vert\psi\rangle\langle\psi\vert$. 

The link with the measurement results is given by the Born statistical interpretation, where the expectation value $E_{\rho}(A)=\langle\psi\vert {A}\psi\rangle$ is interpreted as the value one obtains when one realizes several measurements of the system by means of an apparatus $\mathcal{A}$ (which corresponds to ${A}$\,) and estimates the statistical mean of the measurement outcomes.\cite{frequency}

If we assume that $F$ and $R$ are the only objects required to define a physical theory $T$ in the sense that the statistical regularities are not required to be  further explained, then we get what is called a \textit{minimal instrumentalist interpretation} of the theory.\cite{Redhead}  

A step further in the interpretation of the formalism of quantum mechanics in the framework of an empiricist point of view is to connect density operators $\rho$ and self-adjoint operators $A$ to the measurement procedure. That can be done as soon as we realize that $\rho$ and $A$ are labels\cite{Ludwig,Muynk} for preparation procedures and measurement procedures, respectively. \cite{Measurement} In this respect, pages 74--77  in Ref.~10 are illuminating (here, one can also find the connection with the empiricist approach of van Fraassen.\cite{Fraassen} ).

%%%%%%%%%%%%%%%%%%%%%%%%%%%%%%%%%%%
\section{Behind the statistical regularities: a realist interpretation}
\label{II}

It is worth noting that the Born statistical interpretation by itself does not provide a direct correspondence either between $\psi$ and some empirical data or between $A$ and some empirical data. 

At variance with the minimal instrumentalist interpretation, one could assume a realist position towards physical theories and try to explain the statistical regularities predicted by the formalism as a consequence of the nature of an underlying physical reality.\cite{Redhead} Often, such a vision is materialized by the introduction of a model $M$, such that each term in $F$ corresponds to a term in $M$, which in turn has a clear (intuitive) physical meaning in the model.

As an example we can consider the hydrodynamic interpretation of quantum mechanics, developed by Madelung,\cite{Jammer} where the square modulus of the wave function is interpreted as the density of a hydrodynamical flow and the phase is interpreted as the velocity potential of the flow.  The wave function thus finds an \textit{interpretation} in terms of classical concepts (density and velocity of a fluid). The term ``interpretation'' we are using here is clearly different from the one introduced in the previous section since the wave function is not directly connected to the results of the measurement but is instead ``interpreted'' by means of the model.

A second example of a realist interpretation of quantum mechanics is the de Broglie double-solution interpretation.\cite{deBroglie} It is interesting to note that recently the analogies between such an interpretation and a classical system of droplets walking on a liquid surface has been studied experimentally.\cite{Couder}  Although all such approaches are suggested by the formalism, they have some degree of arbitrariness and are not necessarily valid. For example, the hydrodynamic interpretation was found to be problematic\cite{Jammer} even though its formalism is derivable from the formalism of quantum mechanics.

%%%%%%%%%%%%%%%%%%%%%%%%%%%%%%%%%%
\section{Changing the formalism}
\label{III}

A third kind of ``interpretation,'' which is usually inspired by a realistic position towards the physical theories, occurs when the formalism $F$ is replaced by a different formalism $F'$ to which there corresponds a new set of rules of correspondence $R'$. Some examples are the formulations of quantum mechanics suggested by Bohm,\cite{Bohm}  Ghirardi-Rimini-Weber,\cite{Ghirardi} and the phase-space formulation of quantum mechanics \cite{Schroeck,Beneduci1,Beneduci2} (to be distinguished from the Wigner formulation).
In the Bohmian formulation the trajectories of the single particles are determined by Newton\rq{}s second law by adding a non-local quantum potential. G-R-W introduce a stochasticity in  the dynamical evolution of the systems in order to explain the reduction of the wave packet process. 
The phase space formulation, where the observables are described by Positive Operator Valued Measures (POVMs), provides a differential geometric foundation of quantum mechanics and allows a derivation of classical and quantum mechanics in a unique setting. \cite{Beneduci1,Beneduci2}

\section{On the unavoidability  of the interpretation of quantum mechanics}

As we have seen, an empirical theory $T$ is divided into a formalism $F$ and a system of rules $R$ that connect the formalism to the measurement results.  Because $F$ without $R$ would be a mathematical theory, devoid of any empirical content, the introduction of $R$ is unavoidable.

Now, we have several options: (1) stick with the empiricist's interpretation, which in some sense is minimal; (2) opt for a realist approach, which implies the necessity of an interpretation of one (or two) of the concepts of state, observable, and measurement,\cite{States} referring to an existing physical entity (see Sec.~\ref{II}); or (3) try to change the formalism.

In the first case, quantum mechanics (or more generally a physical theory) is only an instrument to make previsions about the physical phenomena. It is not a description of a veiled reality. As a consequence, all the interpretative problems of quantum mechanics\cite{Examples} that are at the origin of the studies about the interpretation of quantum mechanics disappear at once. Thereby, in the instrumentalist view one does not need to resort to the macroscopic nature of the measurement instruments in order to solve the measurement problem because there is no measurement problem! 

The empiricist view seems to be the choice made by Henry \cite{Henry} but not the choice made by Hobson. \cite{Hobson} It is not clear to us if this is the path followed by van Kampen because the sentence ``the entanglement between two electrons is a manifestation of  the wave function" in van Kampen's letter seems to suggest  an interpretation of the wave function as described in Sec.~\ref{II}. Indeed, that is precisely what happens in the letter by Hobson \cite{Hobson} where the expression  ``manifestation of  the wave function" is replaced by the expression ``manifestation of a matter field."  Thus, Hobson \cite{Hobson,Hobson1} is not rejecting the trend of interpreting quantum mechanics but is proposing his personal interpretation (a realist one) of quantum mechanics (the universe is made of real quantum fields\cite{Hobson1})  as more powerful because (in his view) it is not an interpretation! 

Now that we have clarified that all of the authors of these letters are giving a particular interpretation of quantum mechanics, we would like to analyze the third of the possible kinds of interpretation.   As discussed, this view implies a commitment to a non-standard formalism, and this seems to be the main target of the critiques contained in the letters by van Kampen and Hobson.\cite{Kampen,Hobson,Hobson1}  In any event, it is worth mentioning that an interpretation of the third kind is avoidable only if one supposes that the standard formalism of quantum mechanics is the better one or is the only one we need, but such a viewpoint contains some kind of dogmatic assumption. 
 
We used the Bohm interpretation as an example of interpretation committed to a modified formalism. Now, we would like to remark that Bohm's 1952 papers ``should be not taken as his dramatic conversion to a deterministic, mechanical viewpoint. He was merely trying to show that an alternative [to the Copenhagen interpretation] that attributed properties to an underlying reality was possible'' (Ref. 22, page 7). ``The theory was not proposing the existence of a classical `rock-like' particle, but rather a new kind of entity which is quite different from a classical particle'' (Ref. 22, page 9).  Therefore, it seems that Bohm's approach was inspired by a realist position toward quantum mechanics; i.e.\ quantum mechanics describes some element of reality. Thus, in a certain sense we could say that Bohm and Hobson are aiming at the same target.

But all of this makes it even less understandable why an interpretation that involves a modification of the formalism should be considered ``scandalous.''  Indeed, such an interpretation is simply a tentative grasping of a piece of reality and it is not so clear why it must be (\textit{a priori}) less successful than standard quantum mechanics in this endeavor. The article by Nikoli\'c\cite{Nikolic} is a clever argument supporting such a modified viewpoint. Indeed, as he argues, in a hypothetical version of the history where the Bohmian formulation is proposed before the Born probabilistic interpretation, the Copenhagen interpretation would probably have not achieved great popularity\cite{Nikolic} and the Bohmian interpretation would have been the standard one. It is curious that the article by Nikoli\'c  is cited by van Kampen as a reminder of the scandal of quantum mechanics. We think it actually supports the thesis that the concept of interpretation is both crucial and unavoidable in quantum mechanics.


\begin{thebibliography}{30}
\bibitem{frequency} Clearly, such an interpretation faces all the problems of a frequency interpretation of probability. Anyway, it is used by physicists in their daily activity and therefore it will be used in the present paper in order to illustrate the fact that a connection between the formalism and the measurement procedures must be established.
\bibitem{Measurement} An experiment can be divided into two different parts: a preparation procedure and a measurement or registration procedure. The first is characterized by a preparation apparatus (like an electron emission grid, cyclotron, laser, filter, etc.) that is used to prepare a microscopic system in which we want to measure a particular observable. For example, a filtering operation ensures that if a beam of electrons passes through the filter, the electrons in the beam must immediately afterward have positions within a restricted range of values. The preparation procedure is described by a particular density operator $\rho$ in a Hilbert space $\mathcal{H}$.  Usually, the expression, ``the system has been prepared in the state $\rho$'' is used. 
(Note that it is possible that different preparation procedures correspond to the same density operator.)
It is worth noting that sometimes the filtering procedure is incorrectly identified with a measurement. The measurement procedure is characterized by a measuring instrument (like a photo diode, bubble chamber, spark chamber, etc.) endowed with a macroscopic pointer whose position after the measurement determines the measurement outcome, perhaps within some interval.
This is mathematically represented by a self-adjoint operator $A$ or by a positive operator-valued measure $F$ in $\mathcal{H}$.
The measurement procedure reproduces the quantum-mechanical expectation value $\textrm{Tr} (A\rho)$ when applied to a sequence of identical copies of the system prepared in the state $\rho$.  For a rigorous and more complete treatment see Refs. 9, 10, and 19.
\bibitem{States}  One may define the state $s$ of a physical system as the set of values $E_s(A)$ for all observables $A$. Or one may define an observable $A$ as the connection between states $s$ and outputs $E_s(A)$.\cite{Schroeck,Ludwig} In this trinity (states, observables, expectation values) one may define any one in terms of the other two. What one may not do, at the risk of being called illogical, is to interpret all three of the trinity at the same time.
\bibitem{Examples} For example, ``What is the physical meaning of the wave function?'', ``What should the physical interpretation of the indeterminacy inequality for position and momentum be?'', etc.
\end{thebibliography}
\end{document}